\documentclass[12pt,a4paper]{article}
\usepackage{mymacros}
\begin{document}
\title{Towards an ``AdS$_\mathbf{1}/$CFT$_\mathbf{0}$'' correspondence from the D$\mathbf{(-1)/}$D$\mathbf{7}$ system?}
\author[a]{Sergio E. Aguilar-Gutierrez,}
\author[b]{Klaas Parmentier,}
\author[a]{and Thomas Van Riet}
\affiliation[a]{Instituut voor Theoretische Fysica, K.U. Leuven,\\ Celestijnenlaan 200D, B-3001 Leuven, Belgium}
\affiliation[b]{Department of Physics, Columbia University,\\ 538 West 120th Street, New York, NY 10027, USA}
\emailAdd{sergio.ernesto.aguilar@gmail.com}
\emailAdd{k.parmentier@columbia.edu}
\emailAdd{thomas.vanriet @kuleuven.be}
\abstract{We argue that a type IIB Euclidean supergravity solution of the form $\mathbb{R}\times S^1\times \mathbb{T}^8$ with imaginary self-dual $F_1$ flux through $\mathbb{R}\times S^1$ belongs to the chain of $\AdS_d\times S^d\times \mathbb{T}^{10-2d}$ vacua with (imaginary) self-dual $F_d$ flux,  where $d\leq 5$. Such vacua come from the near-horizon of D$(d-2)$/D$(8-d)$ branes and are supersymmetric for odd values of $d$. For $d=1$ we speculate that the hallmark  of conformal symmetry for the matrix model dual is a vanishing free energy. The matrix dual was recently constructed by \cite{Billo:2021xzh} by adding matrix interactions coming from strings stretching between the D$(-1)$ and D$7$ branes to the IKKT matrix model. We find that the corresponding supergravity solution indeed has vanishing on-shell action. Specific $F_5$ fluxes need to be switched on as a consequence of (a T-dual version of) the Hanany-Witten effect.}
\maketitle
\section{Motivation: A chain of AdS vacua}
\label{sec:Motivation}
The simplest holographic backgrounds in string/M-theory are of the Freund-Rubin type $\AdS_5\times $X$^{5}$, $\AdS_4\times $X$^7$ and $\AdS_7\times$X$^4$ where X$^n$ is an Einstein manifold of positive curvature with a curvature radius that is of the same order as the AdS part of the space. These backgrounds are solutions to 10d or 11d gravity coupled to respectively a 5-form, 7-form, or 4-form field strength. For the 5-form that is exactly the truncation of 10d supergravity (SUGRA) for which there is no coupling between the field strength and the dilaton, whereas 11d SUGRA has no dilaton. In the presence of a dilaton coupling one needs slightly more work, and the next-to-simplest AdS vacua can be found in the chain:
\begin{equation}\label{chain}
    \AdS_d \times S^d\times \mathbb{T}^{10-2d}\,,
\end{equation}
with $d=2,3,4,5$. The odd $d$ solutions are SUSY-vacua of IIB, where the vacuum for $d=5$ (described earlier) preserves 32 supercharges and is holographically dual to $\mathcal{N}=4$ SYM. The $d=3$ solution preserves 16 supercharges and is dual to the so-named D$1$-D$5$ CFT.  The even $d$ solutions are all non-SUSY IIA solutions and we remain agnostic about their stability, since it is not relevant to our discussion here.  All the above AdS solutions for $d<5$ describe near-horizon geometries of D$(d-2)$-D$(8-d)$ bound states where the D$(8-d)$ stack is wrapped over the $\mathbb{T}^{10-2d}$.  These vacua always feature a stack of certain D-branes together with their magnetic dual cousins, and the fluxes obey an (almost) self-duality relation (in 10d string frame)
\be
F_d = (1 \pm \star_{2d})\mathcal{F}\,,
\ee
with $\mathcal{F}$ some $d$-form. For odd values of $d$ this means the fluxes are (anti)-self-dual. In Appendix \ref{App:chain} we briefly recall how to construct this chain of vacua. In Euclidean signature the almost self-duality condition changes into $F_d = (1 \pm \rmi\star_{2d})\mathcal{F}$, and for odd values of $d$ this implies an imaginary self-duality (ISD) condition: 
\begin{equation}\label{ISD}
    \star_{2d}F_d=\mp \rmi F_d\,,
\end{equation}
It is tempting to fantasize about extrapolating the chain \eqref{chain} towards:
\begin{equation}
``\AdS_1 \times S^1"\times \mathbb{T}^8\,,
\end{equation}
which one expects to be supersymmetric as the near-horizon of a D$(-1)$/D$7$ stack with the Euclidean D$7$-branes wrapped over the $\mathbb{T}^8$. The aim of this paper is to push holography to its boundary and construct those ``$\AdS_1$'' vacua and their duals.  We put ``AdS$_1$'' between quotation marks since $1$-dimensional manifolds have no curvature. One way to still define a cosmological constant, or AdS length $L$, would be to use the length of the $S^1$ since for all the Freund-Rubin vacua the AdS$_n$ length scale equals the length scale of the $n$-sphere and so this allows a natural extension to $n=1$.  

To the best of our knowledge, there is no known SUGRA solution describing the  
D$(-1)$/D$7$ intersection.  We believe this is due to two subtleties which will be discussed below in some detail. The first subtlety is about reality conditions for $p$-form gauge fields in Euclidean SUGRA; to describe D$(-1)$-brane solutions one needs to flip the sign of the kinetic term of the $F_1$ field strength, whereas for describing Euclidean D$7$ branes the sign should be the conventional one of the Lorentzian theory. This seems a hurdle for describing a D$(-1)$/D$7$ bound state. The second subtlety is about the timelike T-dual to the Hanany-Witten effect \cite{Hanany:1996ie}; one can think of a D$(-1)$/D$7$ intersection as being timelike T-dual to a D$0$/D$8$ bound state. But the latter bound state does not exist by itself since a fundamental string stretching between the D0 and the D8 brane is necessarily created. One would expect a similar, T-dual, effect for a D$(-1)$/D$7$ bound state and this has been described in a worldsheet context in \cite{Billo:2021xzh}. In the next section we investigate the first issue of conflicting reality conditions. Then, in section  \ref{sec:branesolutions}, we recall the D$(-1)$ and D$7$ brane solutions of Euclidean IIB supergravity and explain the obstacle for a simple bound state solution. In section \ref{sec:T8sol} we construct simple Euclidean vacua which we will argue to correspond to our sought-for $\AdS_1\times S^1$ vacua, whereas in section \ref{sec:A dual conformal} we suggest that the matrix dual description can be found in \cite{Billo:2021xzh}. We conclude with many open problems for further research in section \ref{sec:conclusions}.

\section{Reality conditions in Euclidean supergravity}
It is known that supersymmetry in spacetimes with even signature has ambiguities related to the sign of kinetic terms, even in the case of maximal supersymmetry such as type II SUGRA \cite{Hull:1998vg, Bergshoeff:2000qu}. The signs need to be fixed by physical principles. For instance, for Euclidean AdS SUGRA theories with Euclidean CFT duals, the holographic correspondence can fix these signs \cite{Hertog:2017owm, Bobev:2018ugk}.

In the case of type II SUGRA we can see these issues arise in a simple manner. Consider Wick-rotating the Lorentzian theory to Euclidean signature by the usual procedure in which $t=-\rmi t_E$ with $t_E$ Euclidean time. Since the extremal SUGRA $p$-brane solutions are the fundamental solitons describing D-branes, one should study what the effect of Wick rotation is on these solutions. As they are static, the metric is trivially Wick-rotated. The dilaton also depends on a spatial coordinate and so remains unchanged. The gauge field however can be more involved. In case of an electric $p$-brane solution the gauge-field $A_{p+1}$ has a component along time since
\be
A_{p+1}\sim \epsilon_{p+1}
\ee
with $\epsilon_{p+1}$ the worldvolume tensor along the brane surface. Hence we see that the gauge field gets an extra factor $\rmi$ upon Wick rotation:
\be
A_{p+1}\rightarrow \rmi A_{p+1}\,.
\ee
An imaginary gauge field is effectively described as a real gauge field with wrong sign kinetic term.

If we instead describe a $p$-brane as a magnetic object then there seems to be no Wick rotation of the dual $A_{7-p}$ gauge field since the magnetic Ansatz is
\be
F_{8-p}\sim \epsilon_{8-p}\,,
\ee
 with $\epsilon_{8-p}$ the worldvolume tensor of the unit $(8-p)$-sphere of the transversal space to the brane in radial coordinates. So, confusingly it seems that ``wrong'' sign kinetic terms only appear if we wish to describe branes in the electric frame. Clearly this becomes puzzling whenever we wish to describe a brane and its magnetic dual at the same time. This is exactly the situation which describes the chain of AdS vacua in \eqref{chain}, since these vacua are near-horizons of bound states between stacks of D-branes and their magnetic dual partners. The ISD property of the flux \eqref{ISD} makes it apparent that the gauge fields will have real and imaginary parts.

Aside from the confusion of naively needing both signs of the kinetic term for $p$-form field strengths, one can worry that wrong sign kinetic terms lead to ill-defined path integrals as the action is unbounded. This is not the case on the condition one uses proper boundary conditions and hence proper total derivatives. This is well-known and is for instance discussed  in \cite{Burgess:1989da}. We recall the argument here since it also clarifies in a different way why electric $p$-branes require opposite signs for the kinetic term compared to magnetic branes. 

Consider gravity in $D$ spacetime dimensions containing $n_t$ timelike dimensions coupled to an $(n-1)$-form $A_{n-1}$ with field strength $F_n=\rmd A_{n-1}$,
\be
S[A] = \int \qty(\star R-\tfrac{1}{2}\star \rmd \phi\wedge \rmd \phi -\tfrac{1}{2}\rme^{a\phi}\star F_{n} \wedge F_{n}) \,.
\ee 
The dilaton coupling $\rme^{a\phi}$ to the gauge field is parametrized by some real number $a$, which we do not specify.
The gravity kinetic term is only kept as a reference to our convention for the overall sign of the action\footnote{We define $\star$ by:
$ \star (\rmd x^{\mu_1}\wedge\dots \rmd x^{\mu_p}\wedge\dots )=\tfrac{\sqrt{\abs{g}}}{(D-p)!}{\varepsilon_{\nu_1\dots\nu_{D-p}}}^{\mu_1\dots\mu_p}\rmd x^{\nu_1}\wedge\dots\wedge\rmd x^{\nu_{D-p}}$
and we set $\varepsilon_{01,\ldots, D}=1$.}.

Imagine a situation in which we want to describe path integrals with boundary conditions involving fixed magnetic charges. This means we need to keep $F$ fixed at the boundary. To do this, we can forget that $F$ comes from $A$ and treat $F$ as fundamental. This requires a Lagrange multiplier $B$ that reminds us that $F$ is closed and hence locally exact:
\be\label{SFB}
S[F, B] = \int (\star R-\tfrac{1}{2}\star \rmd \phi\wedge \rmd \phi -\tfrac{1}{2} \rme^{a\phi}\star F_{n} \wedge F_{n} +  B_{D-n-1}\wedge \rmd F_n)\,. 
\ee 
Note that $S[F,B]$ contains an extra term compared with $S[A]$. The EOM for $B$ indeed closes $F$, and we define 
\begin{equation}
    \rmd B_{D-n-1} \equiv G_{D-n}.
\end{equation}
We will immediately see that $G_{D-n}$ is the Hodge dual field strength to $F_{n}$. After integration by parts we can rewrite \eqref{SFB}:
\be
S[F, B] = \int (\star R-\tfrac{1}{2}\star \rmd \phi\wedge \rmd \phi -\tfrac{1}{2}\rme^{a\phi}\star F_n \wedge F_n + (-1)^{D-n}G_{D-n}\wedge F_n + T)\,,  
\ee
where $T$ is the total derivative $T=(-1)^{D-n-1}\rmd (B\wedge F)$. The EOM for $F$ gives:
\be
\rme^{a\phi}\star F_n = (-1)^{(D-n)}G_{D-n}\qquad \rightarrow\qquad
F_n =(-1)^{(D-n)(n+1)+n_t}\rme^{-a\phi}\star G_{D-n}\,.
\ee
If we fill that back into the action we get
\be\label{SFB2}
S[F,B] = \int (\star R-\tfrac{1}{2}\star \rmd \phi\wedge \rmd \phi + \tfrac{1}{2}(-1)^{n_t}\rme^{-a\phi}\star G_{D-n}\wedge G_{D-n} + T)\,.
\ee
This result implies that only for an odd number of time dimensions, the kinetic term is again of the usual sign\footnote{If instead we would substitute $F$ and $\star F$ with its expression in terms of $G$ in the action $S[A]$, it formally becomes $S[A] = \int\star R -\tfrac{1}{2}(-1)^{t}\star G\wedge G$.
Now we get the opposite: with an even number of time dimensions the action has ``the wrong sign''.}. The upshot is that Hodge duality needs to be done carefully, using the Lagrange multiplier $B$. If we wish to use path integrals with boundary conditions that have fixed charge then $S_{F,B}$ is our fundamental action and this fundamental action, when written in its Hodge dual form \eqref{SFB2}, seems to have a wrong sign kinetic term in Euclidean signature naively making the action unbounded from below. But there is a total derivative $T$ in \eqref{SFB2} which in turn leads to \eqref{SFB} and the latter action does not have an unbounded kinetic term. In other words, the total derivative, which does not affect classical equations of motion, nonetheless reflects a choice of boundary conditions and makes the action bounded from below\footnote{In 1d field theory this is simply related to a Legendre transformation.}. For the 
D$(-1)$ instanton solution the term $T$ is the only term contributing to the real part of the on-shell action, reproducing the well-known on-shell action $ng_s^{-1}$ with $n$ the number of instantons and $g_s$ the string coupling.

This discussion about Hodge duality is consistent with our previous reality conditions for gauge fields in order to describe Euclidean $p$-branes. For instance, a Euclidean D$0$ brane is electrically charged under $C_1$, and to fluctuate the gauge field at fixed charge we need a Dirichlet boundary condition on $\star F_2\sim F_8$ and hence we need the description \eqref{SFB} for $F_8$ leading to the ``wrong sign'' action \eqref{SFB2} for $G=F_2$. The opposite reasoning holds for describing Euclidean D$6$ branes. 

This begs the question: what about bound states of branes and their magnetic duals? For instance the Euclidean D$0$/D$6$ bound state. One could work with a doubled formalism (democratic formulation \cite{Bergshoeff:2001pv}), which effectively amounts to having gauge fields with a real and an imaginary part. Given this, one could worry that the metric in this case will generically be complex since the EM tensor of the gauge field will be complex. This however need not occur. The Lorentzian solutions describing brane bound states are static and so Wick rotation preserves reality of the metric. It is easy to check that this is due to cancellations in the EM tensor such that it remains real. For instance for our near-horizon AdS solutions \eqref{chain} we have imaginary self-dual fluxes $\star_{2p} F_p=\rmi F_p$ in such a way that the EM tensor is real. This does not mean that complex metrics have to be disregarded, they just do not occur for the Euclidean bound states in \eqref{chain}. The situation may however be different for the D$(-1)$/D$7$ bound state since the trace-reversed Einstein equations for a 1-form field strength $F_1$ are:
\be\label{F1F1}
R_{ab} = \frac{1}{2}(F_1)_{a}(F_1)_b\,.
\ee
If $F_1 = F_1^R +\rmi F_1^I$, the imaginary cross terms do not cancel, whereas for higher form fields there is an extra index contraction which can be zero if fluxes point in the right directions. 
We will make this explicit next by constructing an example of a  D$(-1)$/D$7$ bound state. Interestingly, our sought-for $``\AdS_1 \times S^1"\times \mathbb{T}^8$ vacuum will nonetheless feature a real metric since $F_5$ flux comes to the rescue, in line with the worldsheet description of \cite{Billo:2021xzh}. This will be shown  in section \ref{sec:T8sol}.

\section{Euclidean brane solutions}\label{sec:branesolutions}
In what follows we describe the Euclidean D$7$ solution and the D-instanton. We also explain why the naive D$(-1)$/D$7$ bound state does not exist as a consequence of (the timelike T-dual to) the Hanany-Witten effect. 

The relevant part of the bosonic Lagrangian of Euclidean IIB SUGRA \cite{Green:1981yb, Schwarz:1983qr} needed for describing the well-known instanton \cite{Gibbons:1995vg} and 7-brane \cite{Greene:1989ya} solutions is:
\begin{equation}
    \frac{\mathcal{L}}{\sqrt{g}} = R - \frac{1}{2} (\partial \phi)^2 - \frac{1}{2} \rme^{2\phi}(\partial \chi)^2\,,
\end{equation}
with $\phi$ the 10d dilaton and $\chi$, the RR zero-form potential with field strength $F_1=\d\chi$. 

The equations of motion  are
\begin{align}\label{eq:axdilreal}
&\nabla_\mu(\rme^{2\phi}\partial^\mu \chi) =0\,,\\\label{eq:axdilreal2}
&\nabla^2 \phi - \rme^{2\phi}(\partial \chi)^2 =0 \,,\\
& R_{\mu \nu} = \tfrac12 (\partial_\mu \phi \partial_\nu \phi + \rme^{2\phi}\partial_\mu \chi \partial_\nu \chi)\,.\label{eq:einst}
\end{align}
For either Euclidean D$7$ branes or D-instantons, it suffices to take the metric Ansatz, in Einstein frame, 
\begin{equation}\label{eq:ansatz}
    \rmd s^2_{E} = \sum_i\rmd x^2_i + H_7(\rmd x^2 + \rmd y^2)\,,
\end{equation}
with $x_i$ denoting the $0, 1,..., 7$ directions.

\subsubsection*{D$(-1)$ brane}
Let us consider the Euclidean type IIB theory. 
A solution can be obtained from the above by taking \cite{Gibbons:1995vg} 
\begin{equation}
    H_7 = 1, \quad \rme^{\phi} = H_{-1}, \quad \chi = \frac{-\rmi}{H_{-1}}, \quad \delta^{\mu\nu}\partial_\mu\partial_\nu H_{-1} = 0\,,
\end{equation}
so that the solution is determined in terms of one harmonic function $H_{-1}$. The string frame metric is flat and obtained upon multiplying the metric by $\rme^{-\phi/2} = H^{-1/2}_{-1}$. 

\subsubsection*{D$7$ brane}

 A compact way to describe the (Euclidean) D$7$ brane relies on the complex combinations $\tau= \chi + \rmi \rme^{-\phi}$, and $z=x + \rmi y$. If we assume $\chi, \phi$ only depend on $x,y$ then the equations \eqref{eq:axdilreal} and \eqref{eq:axdilreal2} can be written as \cite{Greene:1989ya}:
\begin{equation}\label{eq:axdil}
   \partial \bar{\partial}\tau - 2 \frac{\partial \tau \bar{\partial} \tau}{\tau -\bar{\tau}} = 0\,.
\end{equation}
The axion-dilaton equation \eqref{eq:axdil} can be easily solved by taking $\tau(z)$ or $\tau(\bar{z})$ purely (anti-) holomorphic. This means $(\partial_x + \rmi \partial_y)\tau = 0$, which can be expressed with $x^i=\{x,\:y\}$ as
\begin{equation}
    \partial_i \chi = \epsilon_{ij} \partial_j \rme^{-\phi}.
\end{equation}
In particular $\chi$ and $\rme^{-\phi}$ are harmonic.
The Einstein equations for our Ansatz 
are solved by taking 
\begin{equation}
    \rme^{-\phi} = H_7,
\label{eq:dil$D7$}\end{equation}
and then, automatically, $H_7$ must be harmonic.\footnote{The Einstein equation allows for terms $f(z)f(\bar{z})$ multiplying $H_7$ in the metric. These correspond to coordinate re-definitions, but one can take them singular near the brane, such that the asymptotic boundary conditions are changed to correspond with the expected deficit angle from energetic considerations and the well-known fact that one needs 24 $D7$-branes for the $(x,y)$-space to become $S^2$. Typically one starts by imposing the condition of $\SL(2,\mathbb{Z})$-invariance for the axion-dilaton and takes as a solution $j(\tau) = z$ where $j$ is the modular invariant Klein-j-function, having monodromy $1$ around $z=\infty$, meaning that it describes a brane of unit charge, and then one adds other functions to give the deficit angle. This is detailed in \cite{Bergshoeff:2006jj}.}

\subsubsection*{No D$(-1)$/D$7$ bound state from harmonic superposition}

The usual ``harmonic superposition rules'' of \cite{Tseytlin:1996bh} and \cite{Bergshoeff:1996rn} may be naively expected to give the  D$(-1)$/D$7$ bound state solution where the D$(-1)$ branes are smeared over the D$7$ surface. However, it can be verified that this fails \cite{Matthias}. If it would have worked than one could expect a solution with a constant dilaton since the superposition rule entails:
\begin{equation}\label{eq:harmruledil}
    \rme^\phi = H_{-1} H^{-1}_7.
\end{equation}
If we assume that both harmonic function $H_{-1}$ and $H_7$ depend on the mutually transverse coordinates, then we can consider the special case where they are equal. In that case the dilaton is a constant. To illustrate the problem, assume that indeed a constant dilaton is possible as the near-horizon of the would-be intersection. Then, equations \eqref{eq:axdilreal}-\eqref{eq:axdilreal2}  imply that $\chi$ is constant as well, which in turn implies, through \eqref{eq:einst}, that the metric is flat. So we only find the empty $\mathbb{R}^{10}$ vacuum solution. Allowing for a non-constant dilaton in \eqref{eq:harmruledil} does not alleviate the issue, as was checked in \cite{Matthias}.

The failure of the harmonic superposition rule is not some simple consequence of the different reality conditions for the RR axion, but a true obstacle that cannot be fixed in any obvious way by being more liberal in terms of reality conditions. In fact, one can understand a deeper underlying reason for this failure by recalling another example in the literature where the harmonic superposition rule fails; the D0/D8 bound state. A solution can only be found by adding a fundamental string that stretches between the D0 and the D8 \cite{Massar:1999sb, Janssen:1999sa, Imamura:2001cr, Bergshoeff:2003sy} and this is nothing but the Hanany-Witten effect \cite{Hanany:1996ie} that is being captured by SUGRA. Interestingly the T-dual of the D0/D8 bound state would exactly be our sought-for D$(-1)$/D$7$ bound state and we can understand the failure of the harmonic superposition rule from T-duality. If we T-dualize the D0/D8/F1 solutions of \cite{Massar:1999sb, Janssen:1999sa, Imamura:2001cr, Bergshoeff:2003sy}, it would lead us to D$(-1)$/D$7$/W bound state solutions of Euclidean IIB where W is a Euclidean ``wave'' from T-dualising the fundamental string. This solution is described in Appendix \ref{app:wave}. We have found that the corresponding complex metric does not allow a near-horizon limit. For this reason, we will not pursue this direction any further. A more careful string theoretic analysis of the D$(-1)$/D$7$ system shows that the T-dual version of the Hanany-Witten effect\footnote{See Footnote 8 of \cite{Billo:2021xzh}, and \cite{Billo:1998vr} for related discussion.} is actually manifested in a magnetic worldvolume flux on half of the D$7$ branes. This phenomenon motivates us to directly solve equations of motion and Killing spinor equations for Euclidean solutions with imaginary self-dual $F_1$ flux in a 2d space transverse to a $\mathbb{T}^8$. So, we enforce a $\mathbb{T}^8$ factor and, in the spirit of the worldsheet analysis \cite{Billo:2021xzh}, add $F_5$ fluxes, which will indeed cure the problem and allow a simple vacuum solution with ISD $F_1$ flux.

Before we describe our pure flux solution we remark that there are other examples for which the harmonic superposition rule fails to provide a solution. Consider for instance the following intersection of Euclidean D$3$ branes in $\mathbb{R}^{10}$:
\be\label{eq:$D3$$D3$fail}
\begin{aligned}
& \times \times \times \times - - - - - - \\
&  - - - - \times \times \times \times - -\\
\end{aligned}
\ee
Despite the fact that the number of dimensions for which a brane worldvolume direction is orthogonal to the other brane is a multiple of 4, the harmonic product rule Ansatz does not solve the equations of motion. This is different from, for example, the following Euclidean D1 intersection in $\mathbb{R}^{10}$
\be
\begin{aligned}
& \times \times - - - - - - - - \\
&  - - \times \times - - - - - -\\
\end{aligned}
\ee
It is likely that all examples for which the harmonic product Ansatz fails are related to the D$(-1)$/D$7$ intersection by T-duality (and hence to the D$0$/D$8$ intersection). Indeed, the D$3$/D$3$ intersection in \eqref{eq:$D3$$D3$fail} can be obtained by 4 T-duality operations along the D$7$ worldvolume. It would be interesting to find the extra SUGRA fields needed to turn it into a solution, as was done for the D$0$/D$8$ intersection displaying the Hanany-Witten effect.

\section{ \texorpdfstring{``$\AdS_1 \times S^1"\times \mathbb{T}^8$}{} solutions}\label{sec:T8sol}

\subsection{Instanton bound states and fluxes}

It is well known \cite{Douglas:1995bn} that bound states of branes can often be described as field theory solitons on the worldvolume of one of the higher-dimensional branes. For D$(-1)$/D$3$ and D$(-1)$/D$7$ bound states, the D$(-1)$ can be viewed as a YM instanton on the D$3$ or D$7$ brane. The backreaction of such a YM instanton indeed behaves as the SUGRA solution corresponding to D$(-1)$/D$3$ and D$(-1)$/D$7$ bound states. We briefly recall this below, as it will be the inspiration for the Ansatz leading to  ``$\AdS_1 \times $S$^1"\times \mathbb{T}^8$ solutions.

\subsubsection*{D$(-1)$/D$3$}
Consider $N$ D$3$ branes on which we have a U($N$) gauge theory. If we stick to the gluon sector, then the Euclidean action in 10d string frame contains
\begin{equation}\label{$D3$action}
S \supset \frac{1}{g_s}\int \rmd x^4 \sqrt{g}\qty(N + \text{Tr} (F^2)) + \rmi C_0 \int \text{Tr}(F\wedge F).     
\end{equation}
The constant term is the D$3$ tension. One can then consider a classical instanton configuration for which
\begin{equation}
\star F = \pm F\,,    
\end{equation}
and
\begin{equation}\label{BPS}
\text{Tr}(F^2) =\pm \text{Tr}(F\wedge F) \sim k    \frac{\rho^4}{\left((\vec{x}-\vec{x}_0)^2 + \rho^2\right)^4}\,,
\end{equation}
with $k$ the instanton charge
\be
k\sim \int \Tr(F\wedge F)\,,
\ee
$\rho$ is the thickness and $\vec{x}_0$ the position. From \eqref{$D3$action} we see that these gauge fields source the dilaton and the axion exactly like a D$(-1)$ brane does. For the axion that is obvious: the WZ term literately becomes the WZ term of a D$(-1)$ brane with charge $k$. For the dilaton this is most easily seen by going to Einstein frame, where the D$3$ brane action becomes
\begin{equation}
S \supset  \int \rmd x^4 \sqrt{g}\left( N + \frac{1}{g_s} \text{Tr}(F^2) \right).     
\end{equation}
So the tension does not scale with $g_s$, but the gluon term scales as $g_s^{-1}$ and therefore the YM instanton on the D$3$ backreacts like the D$(-1)$ brane, whose tension in Einstein frame indeed goes like $g_s^{-1}$ (same as in string frame). The BPS condition \eqref{BPS} shows that the D$(-1)$ brane tension equals its charge, as expected. The corresponding supergravity solution can be found in the usual way by multiplying harmonic functions in a precise way and was described in \cite{Liu:1999fc}.

\subsubsection*{D$(-1)$/D$7$}
The story is similar here, see for instance \cite{Billo:2021xzh}. A class of 8d YM instantons is defined by the relation
\begin{equation}
\star(F\wedge F) = F\wedge F\,,
\end{equation}
with the instanton charge $k$ going like:
\be
k\sim \text{Tr} (F\wedge F\wedge F\wedge F)\,.
\ee
Now the YM action on the D$7$ is different:
\begin{equation}\label{$D7$action}
S \supset \frac{1}{g_s}\int \rmd x^8 \sqrt{g} \left(N + \text{Tr} (F^2) + \text{Tr}(t_8 F^4)\right) + \rmi C_0 \int \Tr(F\wedge F\wedge F\wedge F).
\end{equation}
The definition of the $t_8$ tensor appears in \cite{Billo:2021xzh}. Upon moving to Einstein frame we have
\begin{equation}\label{$D7$actionEinstein}
S \supset \int \rmd x^8 \sqrt{g}\left( g_sN + \text{Tr} (F^2) + \frac{1}{g_s} \text{Tr}(t_8 F^4)\right) .
\end{equation}
So we see that the $t_8$ term and the WZ term indeed mimic a D$(-1)$ brane. But now Tr$(F^2)$ does not. It seems to backreact like a D$3$ brane.  Also the WZ term of a D$3$ brane is there since we have also a term in the D$7$ brane action of the form:
\be
C_4\wedge F\wedge F\,.
\ee
This is indeed related to the Hanany-Witten effect as explained in \cite{Billo:2021xzh} where it is observed that this effect implies that we require a specific set of ``background'' worldvolume flux on the D$7$ brane in the presence of D-instantons. Since these worldvolume fluxes behave like D$3$ branes, it strongly suggests that the holographic vacuum solution features both ISD $F_1$ flux and $F_5$ flux with $4$ feet along the $\mathbb{T}^8$ direction. We will indeed see that this is exactly the ingredient that allows a Euclidean $\AdS_1\times S^1\times\mathbb{T}^8$ solution.

\subsection{The \texorpdfstring{$\AdS_1\times S^1\times\mathbb{T}^8$}{} solution}\label{sec:The solution}

Below, we construct the vacuum solution which we should think of as the near-horizon of D$7$/D$3$/D$(-1)$. This means we need $F_5$ fluxes, which are imaginary self-dual in Euclidean signature.  The D$7$ branes, now dissolved into fluxes, are wrapping the 8-torus. The D$3$ branes are located inside the D$7$ brane worldvolume as specified in (\ref{eq:F1 complete}) and the D$(-1)$ branes are considered as smeared over the 8-torus. 

The 10d metric in string frame is flat and written as
\begin{equation}
\rmd s^2 = L_y^2\rmd y^2 + L_x^2 \rmd x^2   +  \sum_{i=1}^8 L_i^2(\rmd \theta^i)^2\,.
\end{equation}
Let us take the $\theta^i$ and $x$ to have periodicity $2\pi$. Whether $y$ is a line or a circle we leave open at the moment. The difference could be related to the difference between Euclidean ``$\AdS_1$'' or Euclidean ``dS$_1$''. In case $y$ is non-compact we can always absorb $L_y$ into the coordinate $y$, making it pure gauge. For convenience we kept $L_y$ throughout. What distinguishes the $y$ direction as Euclidean time is that it is the direction with imaginary $F_1$ flux.  Hence we write:
\be
F_1=\alpha \rmd x + \rmi\beta \rmd y
\ee
with $\alpha$ and $\beta$ real. 
In a doubled formalism with real fields we would instead write:
\be
F_1=\alpha \rmd x \,, \qquad F_9 = \beta \star_{10}  \rmd y\,.\label{eq:F1 complete}
\ee
We assume that the dilaton is constant. 

Note that strictly speaking we do not need to assume any topology for finding a bulk solution, but we consider $x$ and the $\theta$ directions to be circles in order to generalise our chain of $\AdS_d\times S^d\times\mathbb{T}^{10-2d}$ vacua. Upon quantisation of fluxes the topology does matter.

We can take many possible $F_5$ fluxes consistent with the many choices to put SUSY intersecting D$3$-branes inside D$7$ (and hence the 8-torus):  
\be
\begin{aligned}
& \times \times \times \times - - - - \\
&  - - - - \times \times \times \times \\
& - - \times \times \times \times - -\\
& \times \times - - - - \times \times\\
& - - \times \times - - \times \times\\
&\times \times - - \times \times - -
\end{aligned}\label{eq:inter}
\ee
The corresponding fluxes are obtained by defining a $C_4$ profile along the crosses. The expectation is that for every such line of fluxes one halves the SUSY. Since $F_5$ flux is ISD automatically, fluxes for a single D$3$ brane always come with its magnetic dual. The first line then comes together with the second, the third with the fourth, and the fifth with the sixth. So one expects to be able to reduce the SUSY by a factor of $1/2$, $1/4$ or $1/8$.  We thus arrive at the following $F_5$ Ansatz:
\be
\begin{aligned}
F_5&=(1 - \rmi \star) \mathcal {F}\,,\\\label{eq:F5 complete}
\mathcal{F} = \rmd \theta^{1234}\wedge  p_1 &+\rmd \theta^{1256}\wedge p_2+\rmd \theta^{1278}\wedge p_3
\,. \nonumber
\end{aligned}
\ee
We introduced the notation:
\be
p_a= \gamma_a \rmd x + \delta_a \rmd y,\quad \rmd\theta^{ijkl}=\rmd\theta^i\wedge\rmd\theta^j\wedge\rmd\theta^k\wedge\rmd\theta^l
\ee
for $a=1,\,2,\,3$. Assuming a compact topology, we need to quantise the numbers $\alpha, \beta, \gamma_a, \delta_a$ and this will involve all radii $L_x$, $L_y$, $L_i$.  The more $F_5$ directions we switch on, the more moduli of the 8-torus will get stabilised. In the solution we present next we take the simplest fluxes, only featuring $p_1$. In Appendix \ref{app:gen} we present the general result. 

We now work through all the equations of motion of Euclidean IIB SUGRA and analyse which conditions are needed to solve the equations of motion in Einstein frame. Afterwards we will discuss the supersymmetry of the solution.

\begin{itemize}
\item The ISD choice of the $F_1$ flux implies that satisfying the $F_1$ Bianchi identity means one satisfies the $F_1$ equation of motion $\rmd\star F_1=0$. The same applies to the $F_5$ equation of motion. So all flux equations are satisfied. 
\item The dilaton equation of motion is also satisfied. This is easiest to see in Einstein frame, where it reads $\nabla^2\phi =\rme^{2\phi}F_1^2$. A constant dilaton is only possible when $F_1^2=0$ which implies the constraint:
\be\label{constraint1}
\alpha L_y=\pm \beta L_x\,.
\ee
In other words: the $F_1$ flux is (anti-)ISD along the ``$\AdS_1 \times S^1"$ direction.
\item The Einstein equation is solved on the condition
\be
\tfrac{1}{2} \rme^{2\phi}\,(F_1)_a (F_1)_b + \tfrac{1}{4}(F_5)^2_{ab}=0\,, 
\ee
with $(F_5)^2_{ab}=\frac{1}{4!}F_{a\mu_1\dots\mu_4}F_{b}^{\mu_1\dots\mu_4}$. Let us verify this for all directions. In the directions of the 8-torus we find that $R_{\theta_i\theta_j}=0$ for $i\neq j$ is automatically satisfied. Meanwhile, $R_{\theta_i\theta_i}=0$ leads to:
\begin{align}
L_x\delta_1&=\pm \rmi L_y\gamma_1\,. \label{constraint2}
\end{align}
In the $x, y$ directions we find :
\begin{align}
& R_{xx} = 0\,:\qquad \alpha^2 +\frac{\gamma_1^2}{L_{1234}^2}=0\,,\label{constraint3}\\
& R_{yy} = 0\,:\qquad \beta^2 -\frac{\delta_1^2}{L_{1234}^2}=0 \label{constraint4}\,,\\
& R_{xy} = 0\,:\qquad \rmi\alpha\beta + \frac{\gamma_1\delta_1}{L_{1234}^2} =0 \label{constraint5}\,,
\end{align}
where $L_{ijkl}=L_iL_jL_kL_l$.
The second equation \eqref{constraint4} entails the same constraint as the first \eqref{constraint3}, once we use the earlier constraints \eqref{constraint1} and \eqref{constraint2}. The same is true for \eqref{constraint5} \emph{if} we chose the same signs in constraints \eqref{constraint1} and \eqref{constraint2}. 

\end{itemize}

To summarize, we have found the following solution to Euclidean IIB SUGRA in string frame:
\begin{align}
& \d s^2 = L_x^2 \rmd x^2   + L_y^2 \rmd y^2 +  L_1^2(\rmd\theta^1)^2+\ldots+ L_8^2(\rmd\theta^8)^2\,,\label{eq:ENSL1}\\
& F_1=\alpha \rmd x + \rmi\beta \rmd y\,,\\
& F_5 =  (1-\rmi\star)\d\theta^1\wedge\d\theta^2\wedge\d\theta^3\wedge\d\theta^4\wedge (\gamma_1 \d x +\delta_1 \d y)\,,\label{eq:ENSL2}\end{align}
where
\begin{align}
& \alpha L_y=\pm \beta L_x \label{c1}\,,\\
& L_x\delta_1=\pm \rmi L_y\gamma_1 \label{c2}\,,\\
& L_{1234}\alpha =\pm \rmi\gamma_1 \label{c3}\,,
\end{align}
with the signs in the first two constraints taken to be the same.

\subsection{Flux quantisation }\label{sec:Flux quantisation}

Quantisation of the $F_1, F_9$ fluxes implies:
\begin{align}\label{eq:Qalphabeta}
& 2\pi \alpha = N_7\,,\\
& 2\pi \beta =(\alpha')^4 N_{-1}\frac{L_y}{L_x}\frac{1}{\text{Vol}(\mathbb{T}^8)}\,
\end{align}
where $\text{Vol}(\mathbb{T}^8)=L_{1}\dots L_8$. Note that only the combination $\beta/L_y$ appears in the quantisation condition, just like in the constraints from the equations of motion \eqref{constraint1}, consistent with the fact that we can gauge away $L_y$ if Euclidean time is non-compact. 

Flux quantisation of $F_5$ works as follows: 
\begin{align}
\int_{1234x} F_5 &= (2\pi)^5 \gamma_1 \equiv (2\pi)^4(\alpha')^2 N_{3}\,, \\
\int_{5678x} F_5 &= -(2\pi)^5  \rmi\frac{L_x L_{5678}}{L_y L_{1234}}\delta_1 \equiv (2\pi)^4(\alpha')^2 \tilde{N}_{3}\,,
\end{align}
where the $N_{3},\, \tilde{N}_{3},$ are (imaginary) integers. From now on, we will work in string units for which $\alpha'=1$. 

Our constraints (\ref{c1} - \ref{c3}), become then, respectively: 
\begin{align}
& N_7\text{Vol} \mathbb{T}_8 =\pm  N_{-1} \,,\label{qc1}\\
& L_{1234}\tilde{N}_{3}= \pm  L_{5678} N_{3}\,,\label{qc2}\\
& N_7 L_{1234} =\pm \rmi N_{3} \label{qc3}\,.
\end{align}
We notice that the first constraint stabilises the torus volume and the second the volume of the separate 4-tori. 

\subsection{On-shell action}

Our solution of interest has an $\mathbb{R} \times\mathbb{T}^9$ topology. $F_1$ has legs in two directions, one of which is compact (corresponding to $x$ in our notation for the $S^1$), while the other one is a line (corresponding to the Euclidean time version of $\AdS_1$). Other possible topologies for our Ansatz are outside the scope of this paper.

We would like to show that the $\mathbb{R} \times\mathbb{T}^9$ solution fulfills the criterion of having a vanishing on-shell action, in order for the solution to have a dual conformal matrix theory description, which we motivate in Section \ref{sec:A dual conformal}. Given the real line $y$, we need to be careful with boundary terms.  As we explained earlier, when gauge fields have an electric component, it becomes imaginary in Euclidean signature and the proper way to compute the on-shell action then uses Hodge duality. In the electric frame this amounts to including a total derivative, but in the magnetic frame this means one simply works with the Hodge dual part without boundary term. Hence, we will work in the democratic formalism and use $F_9$ instead for the instanton-flux. The string frame action then becomes
\be
S= -\int \sqrt{g}\left(g_s^{-2}\mathcal{R} -\frac{1}{2}F_1^2  -\frac{1}{2}\frac{1}{9!}F_9^2 + \frac{1}{4}\frac{1}{ 5!}(F^E_5)^2  - \frac{1}{4}\frac{1}{ 5!}(F^M_5)^2 \right)\,.
\ee
Note the subtleties with the 5-form, for which one needs to use a trick to compute the on-shell action \cite{Mkrtchyan:2022xrm, Kurlyand:2022vzv}: one splits $F_5$ into electric and magnetic and flips the sign of the electric part\footnote{Here we are following the standard procedure, although one might ask whether the action is still bounded given the reality conditions implied by the constraints (\ref{c1} - \ref{c3}). This issue seems to boil down to the difficulty in finding Euclidean D$3$ intersections of the form \eqref{eq:$D3$$D3$fail}, for which it does not seem possible to unambiguously discern an electric and a magnetic part. In our current setup we used D-instanton flux to decide which is which.}. By imaginary self-duality we know that $(F^E_5)^2=-(F^M_5)^2$ and hence the on-shell action becomes
\be
S= -\int \sqrt{g}\left(g_s^{-2}\mathcal{R} -\frac{1}{2}F_1^2  -\frac{1}{2}\frac{1}{9!}F_9^2 -  \frac{1}{2}\frac{1}{ 5!}(F^M_5)^2 \right)
\ee
Let us now compute this integral. We have 
\begin{align}
& \mathcal{R}=0\,,\\
& F_1^2 =\alpha^2L_x^{-2}\,,\\
&\tfrac{1}{9!}F_9^2 = \beta^2L_y^{-2}\,,\\
&\tfrac{1}{ 5!}(F^M_5)^2 = \gamma_1^2 L_x^{-2}L_{1234}^{-2} - \delta_1^2 L_y^{-2}L_{1234}^{-2}\,.
\end{align}
Using the first two constraint equations \eqref{c1} and \eqref{c2}, the on-shell action becomes
\be
S = \frac{1}{L_x^2} \int \sqrt{g} \left(\alpha^2 +\frac{\gamma_1^2}{L_{1234}^2} \right) \,,
\ee
which nicely adds up to zero thanks to the third constraint \eqref{c3}.

\subsection{Supersymmetry}\label{sec:SUSY part}
In this section we consider the supersymmetry variations in the background (\ref{eq:ENSL1} - \ref{eq:ENSL2}) of Euclidean IIB SUGRA.  It has been argued in the literature that IIA has a real Euclidean formulation whereas Euclidean IIB is not well-understood \cite{Bergshoeff:2007cg,Bergshoeff:2000qu}. Our work will provide more evidence in that direction. To anticipate the subtlety let us recall the fermion SUSY variations in  Lorentzian SUGRA in string frame  \cite{hamilton2016field} 
\begin{align}
   \delta^L_\epsilon \lambda &=\tfrac{1}{2}\qty(\partial_\mu\phi-\rmi\rme^{\phi})\Gamma^\mu {F}_\mu\epsilon\;,\label{eq:SUSY1}\\
   \delta^L_\epsilon\psi_\mu &= D_\mu \epsilon + \tfrac{\rmi}{8} \rme^{\phi}\qty(\Gamma^\nu {F}_\nu+\tfrac{1}{2\cdot5!}\Gamma^{\nu_1\dots\nu_5} {F}_{\nu_1\dots\nu_5}) \Gamma_\mu\epsilon\;.\label{eq:SUSY2}
\end{align}
Notice that we are allowed to exchange $F_5\rightarrow \star F_5$ given the self-duality condition. Now, imagine one rewrites the Lorentzian SUSY variations in terms of the Hodge starred field strengths, i.e. one replaces $F_3$ with $\star F_7$, etc. This is simply a rewriting, and nothing more. After that is done, one could postulate that the Euclidean SUSY variations are simply the ones from this Lorentzian theory but with the Hodge dual notation. In Euclidean signature the self-duality takes a different form: $F_5= \rmi\star F_5$ and then the Euclidean variations will differ once we go back to the notation without Hodge stars. In particular the $F_5$ appearance will have an extra $\rmi$ in front. This means that there is an ambiguity in Euclidean IIB when the equations involve $F_5$, which is reminiscent of twin-supergravities \cite{Anastasiou:2016csv}, namely that there are 2 possible supersymmetry transformations
\begin{align}
\text{Option I\,}:\qquad    \delta^E_\epsilon\psi_\mu &= D_\mu \epsilon + \tfrac{\rmi}{8} \rme^{\phi}\qty(\Gamma^\nu {F}_\nu+\tfrac{1}{2\cdot5!}\Gamma^{\nu_1\dots\nu_5} {F}_{\nu_1\dots\nu_5}) \Gamma_\mu\epsilon,\label{eq:oldSUSY}\\
\text{Option II}:\qquad      \delta^E_\epsilon\psi_\mu &= D_\mu \epsilon + \tfrac{\rmi}{8} \rme^{\phi}\qty(\Gamma^\nu {F}_\nu+\tfrac{\rmi}{2\cdot5!}\Gamma^{\nu_1\dots\nu_5} {F}_{\nu_1\dots\nu_5}) \Gamma_\mu\epsilon\,.\label{eq:novelSUSY}
\end{align}
The well-established example of a stack of D$3$ branes with its $H_5\times S^5$ near-horizon geometry requires the first option, whereas SUSY of our flux solution seems only well-behaved for option II. Let us now verify this. 

For concreteness, we choose the positive signs in (\ref{c1} - \ref{c3}). Imposing that the dilatino variation \eqref{eq:SUSY1} vanishes then leads to a first projector
\begin{equation}
        (\Gamma_{\hat{x}}+\rmi \Gamma_{\hat{y}})\epsilon=0\label{eq:sing proj},
\end{equation}
where the hats denote frame indices, for which $\comm{\Gamma_{\hat{a}}}{\Gamma_{\hat{b}}}=2\delta_{\hat{a}\hat{b}}$.

The gravitino variations with the $F_5$ Ansatz in (\ref{eq:F5 complete}) produces:
\begin{equation}
    \partial_\mu\epsilon+ \tfrac{1}{16}\rme^\phi\Big(2\rmi\tfrac{\beta}{L_y} +K\Big)(\Gamma_{\hat{x}}+\rmi\Gamma_{\hat{y}})\Gamma_\mu \epsilon=0.\label{eq:sec proj1}
\end{equation}
where
\begin{equation}
K=\rmi^z\tfrac{\delta_1}{L_{1234}}\left(\Gamma_{\hat{\theta}^1\hat{\theta}^2\hat{\theta}^3\hat{\theta}^4}+\Gamma_{\hat{\theta}^5\hat{\theta}^6\hat{\theta}^7\hat{\theta}^8}\right)\label{eq:sec proj2}
\end{equation}
after setting $\varepsilon_{xy{\theta}^1,\dots, {\theta}^8}=+1$, and $z$ denotes the index of our options for the gravitino variations: $z=0$ for option I, and $z=1$ for option II. For any value of $z$, (\ref{eq:sec proj1} - \ref{eq:sec proj2}) imply that the KS will not depend on the coordinates $\theta^i$. As for the $x$- and $y$-components, we will use chirality conditions of the type IIB spinor
    \begin{equation}
 \Gamma_{\hat{\theta}^1\dots\hat{\theta}^8\hat{x} \hat{y}}\epsilon=\pm\rmi\epsilon.\label{eq:chirCnd}
    \end{equation}
to study the Killing spinor solutions for each option below.

\subsubsection*{SUSY variations for option I}
Upon choosing the negative sign in (\ref{eq:chirCnd}), (\ref{eq:sec proj1}) further reduce to
\begin{equation}
    \partial_x \epsilon + \rmi\tfrac{\alpha}{4} \rme^\phi \epsilon =0,\qquad\partial_y \epsilon - \tfrac{\beta}{4} \rme^\phi \epsilon =0\,,\label{eq:KSeqold}
\end{equation}
which, given our constant dilaton $\phi$, is solved by
\begin{equation}\label{epssol}
    \epsilon = \rme^{-\frac{\rmi}{4}\rme^\phi(\alpha\,x + \rmi\beta\,y)}\epsilon_0
\end{equation}
where $\epsilon_0$ is a constant spinor. We thus see that this solution naturally identifies the $x$-direction, with real $1$-form flux, as a circle, and the $y$-direction as a line. Note however that \eqref{eq:Qalphabeta} implies that the spinor acquires a phase different from $\pm 1$ when going around the $x$-circle,
\begin{equation}
        \epsilon(x+2\pi,y,\theta_1,\dots,\theta_8)=\epsilon(x,y,\theta_1,\dots,\theta_8)\rme^{\rmi\rme^{\phi}\frac{N_7}{4}}
    \end{equation}
unless one chooses $\rme^{\phi}=2\pi$, which invalidates the supergravity approximation, combined with the quantization condition for the D$7$ charge: $N_7\in2\mathbb{Z}$. In such case, it appears that, at least locally, we have a half-BPS solution, as is indeed expected for near-horizon limit where the branes have effectively dissolved into fluxes. What seems off, besides strong coupling, is that adding flux along more directions in the internal $\mathbb{T}^8$ does not reduce the amount of supersymmetry, as shown in Apendix \ref{app:gen}. This is indeed not what one expects from the analysis of \cite{Billo:2021xzh} in the putative matrix dual to our theory. As such we must explore whether our proposal for the new SUSY variation (option II) can do better.

\subsubsection*{SUSY variations for option II}
Let us explore the case where we choose the positive sign in the chirality condition (\ref{eq:chirCnd}) and  $z=1$ in (\ref{eq:sec proj2}). To solve the Killing spinor equations with a constant spinor, one needs to impose an extra projector, which we take
\begin{equation}
    (1\pm\Gamma_{\hat{\theta}^1\hat{\theta}^2\hat{\theta}^3\hat{\theta}^4})\epsilon=0\,,\label{eq:selectProj1}
\end{equation}
where we choose the sign such that (\ref{eq:selectProj1}) in (\ref{eq:sec proj1} - \ref{eq:sec proj2}) together with (\ref{c1} - \ref{c3}) leads to a constant spinor solution:
\begin{equation}
    \epsilon(x,y,\theta_1,\dots,\theta_8)=\epsilon_0.\label{eq:const spinor sol}
\end{equation}
The solution is now globally-defined independently of the dilaton vacuum expectation value, and it is a 1/4-BPS state; this number is halved to $1/8$ BPS by adding extra $F_5$ flux in the directions shown in (\ref{eq:inter}), as explained in Appendix \ref{app:gen}. However, as shown there, this only holds for an extra constraint on the fields $e^{\phi}, L_{i}$, beyond the equations of motion and we have no clear interpretation for this phenomenon.

\section{A dual conformal matrix model?}\label{sec:A dual conformal}

Consider the following schematic form of a matrix integral
\begin{equation}\label{eqn:Zlambdadef}
  Z[\lambda_i] = \int \rmd M\, \rme^{-V(M,\lambda_i)}\,.
\end{equation}
The set of constants $\lambda_i$ determines the potential and can be viewed as the coupling constants in this theory. While it is hard to properly define a conformally invariant matrix model, one possible definition is to consider rescaling the matrix $M \to \Omega M$ and demanding that the resulting theory stays the same. While this definition may not be ideal it amounts to demanding that the matrix model does not depend on the coupling constants $\lambda_i$ and its partition function is therefore a constant. The precise value of this constant may not have an independent physical definition since in the calculation of any correlation function in the matrix model it will drop out in an appropriate normalization. We therefore arrive at the conclusion that a possible operational definition of conformal matrix models may be taken to be those matrix integrals for which $Z[\lambda_i]=\text{const}=1$. This in turn leads to vanishing free energy $F = \log Z[\lambda_i]=0$. In the context of our discussion this definition is in harmony with the fact that the on-shell action of the supergravity solution described above vanishes. The independence of $Z$ with respect to the marginal couplings is inspired by the independence of the central charge on marginal couplings for CFTs in dimensions two and higher. 

To avoid confusion we emphasize that each of the parameters $\lambda_i$ in \eqref{eqn:Zlambdadef} are the (marginal) couplings of the theory and not external sources. The matrix integral in the presence of sources, $Z[\lambda_i; J]$, should of course depend on the source $J$ and will act as a generating functional for $n$-point functions in the matrix model.

The conformal group on $\mathbb{R}^d$ contains (at least) $SO(1, d+1)$ and so a natural extension to matrix theory implies a $SO(1,1)$ symmetry which would correspond to the scaling symmetry of the $y$-coordinate in the bulk and to the rescaling of the matrices in the dual. 

\textit{Billo et al.} \cite{Billo:2021xzh} proposed a system composed of a stack of $N$ D$7$ branes with magnetic worldvolume flux, adding also $M$ D$7'$ branes without flux, and $k$ D-instantons. They showed that the presence of D-instantons in this system requires a vanishing partition function $Z_{D7'/D(-1)}$ related to the condition dual to the Hanany-Witten effect for the D$0$/D$8$ bound state. To compare this with our supergravity solution, we expect that the overall worldvolume fluxes needed on the D7 are dual to our $F_5$ profiles, since such worldvolume fluxes act more like induced D3 charges and hence become $F_5$ flux in the holographic setup. We indeed saw that we need a fine-tuned $F_5$ flux for supersymmetry and solving the EOM. More specifically, the worldvolume fluxes on half of the D$7$ branes, which are not describing localised instantons, take the form \cite{Billo:2021xzh}
\begin{equation}
2\pi F=\begin{pmatrix} 0 & f_1 & 0 & 0 & 0 & 0 & 0 & 0\\
-f_1 & 0 & 0 & 0 & 0 & 0 & 0 & 0\\
0 & 0 & 0 & f_2 & 0 & 0 & 0 & 0\\
0 & 0 & -f_2 & 0 & 0 & 0 & 0 & 0\\
0 & 0 & 0 & 0 & 0 & f_3 & 0 & 0\\
0 & 0 & 0 & 0 & -f_3 & 0 & 0 & 0\\
0 & 0 & 0 & 0 & 0 & 0 & 0 & f_4\\
0 & 0 & 0 & 0 & 0 & 0 & -f_4 & 0
\end{pmatrix}\mathbbm{1}_{N\times N}
\end{equation}
This describes a particular $U(1)$ flux over the 8-torus. Supersymmetry then requires the $f_I$ to obey: 
\be
\prod_I\frac{1-\rmi f_I}{1+\rmi f_I}=1\,,
\ee
It was observed that in the presence of D$(-1)$/D$7$ strings it is not allowed to put all $f_I$ equal to zero. In other words, this flux is required for consistency once D$(-1)$ branes are considered. 

Out of these 4 constrained numbers $f_I$, the symmetric combinations $f_If_J$ will appear in $\text{Tr}(F\wedge F)$ sourcing the D3 fluxes. Note that indeed we have similarly six possible $F_5$ flux terms labelled by the $\delta_a,\gamma_a$ before. A full correspondence between our flux parameters and the worldvolume fluxes is beyond the scope of the current work since simple attempts have failed. For instance the counting of the number of preserved supercharges does not quite work: whereas reference \cite{Billo:2021xzh} finds that every extra non-zero $f_I$ component halves the amount of SUSY, we only find two options: either $1/4$ or $1/8$ BPS. Furthermore, for the $1/8$ BPS solutions we needed an extra constraint on the field expectation values beyond the constraints from the equations of motion.


After including the D$(-1)$/D$7$ string states, reference \cite{Billo:2021xzh} arrives at an explicit supersymmetric matrix model that we conjecture to be the natural candidate holographic dual to our Euclidean IIB vacuum solution. The matrix model has a gauge group $U(N_7/2)\times U(N_7/2)\times U(N_{-1})\times SU(4)$, where $SU(4)$ is the internal symmetry associated to transformation of the D-branes and rotations in the 4 two-tori inside the eight-torus, indexed by $I$.  Since we leave explicit computations with this matrix model for future work, we refer the reader to the original reference \cite{Billo:2021xzh} for the detailed description.

In order to make a convincing case that the matrix model of \cite{Billo:2021xzh} would be the matrix dual to our vacuum solution we would require that $Z=1$ in absence of external sources. This naively seems in agreement with the explicit expressions for the grand canonical instanton partition function of \cite{Billo:2021xzh} upon putting certain moduli in the origin, but we leave this for future research since such a statement is highly dependent on overall normalisations and a precise treatment tracking down all the appearances of couplings would be required. We hope to come back to this in the future. 

\section{Conclusions and outlook}\label{sec:conclusions}
Let us briefly recapitulate our findings. We started by recalling that there exists a chain of (Euclidean) $\AdS_d\times S^d\times \mathbb{T}^{10-2d}$ vacua in IIA/B supergravity with (imaginary) self-dual $F_d$ flux,  where $d\leq 5$. Such vacua come from the near-horizon of (Euclidean) D$(d-2)$/D$(8-d)$ branes and are supersymmetric for odd values of $d$. We then speculated that there is a natural extension to $d=1$ corresponding to a SUSY bound state of D$(-1)$/D$7$ branes. Such a brane set-up comes with many subtleties related to reality and boundary conditions in Euclidean IIB. On top, this bound state features a timelike T-dual to the Hanany-Witten effect for the D$0$/D$8$-brane system. 

We have not achieved to find localised bound state solutions but we did find the would-be near-horizon vacuum solution of the form $\mathbb{R}\times S^1\times \mathbb{T}^8$ with imaginary self-dual $F_1$-flux through $\mathbb{R}\times S^1$. This required specific $F_5$ flux which we were able to find, inspired by the dual Hanany-Witten effect. A supersymmetry analysis turned out to reveal an interesting ambiguity in the Euclidean IIB supersymmetry variations, which requires further research. We then suggested that the instanton matrix model of \cite{Billo:2021xzh} would be the dual conformal matrix model. We argued that the characteristic $SO(1,1)$ conformal symmetry implies a vanishing free energy which we were able to reproduce in the bulk solution. A more in-depth study of the correspondence between the concrete matrix model of \cite{Billo:2021xzh} and our IIB solution is needed and left for future research since it requires rather involved matrix theory computations. 

Lastly we want to speculate about a broader use of these vacua. Outside the holographic context, the existence of 1d ``vacua'' have been suggested in a cosmological setting \cite{Hertog:2021jyd, Heckman:2018mxl} as seed universes that lead to 3 observably large dimensions after decay. Interestingly the IKKT matrix model \cite{Ishibashi:1996xs}, which has been suggested as a non-perturbative definition of IIB string theory, seems to allow for dynamical solutions in which 3 large classical dimensions emerge \cite{Kim:2011cr}. This approach to early-universe physics captures actual non-trivial string theory effects and vastly differs from the usual 4d EFT-like approaches used to inflation \cite{Brahma:2021tkh, Brahma:2022dsd}. In this regard it is fascinating that our proposal for an ``$\AdS_1$" vacuum leads to a matrix model that is an extension of the IKKT matrix model with extra interactions that make it conformal. It is therefore natural to assume that breaking conformal invariance leads to interesting cosmological dynamics in the bulk that will be described by matrix theory.

\section*{Acknowledgments}
We would like to thank Marco Bill\`o, Eric Bergshoeff, Fridrik Gautason, Miguel Montero, Matthias Vancraeynest and Toine Van Proeyen for useful correspondence and discussions. Special thanks go to Nikolay Bobev for many stimulating discussions and initial collaboration. SEAG, KP and TVR thank Uppsala University for its hospitality during the time when part of this project was carried out. The work of SEAG and TVR is partially supported by the KU Leuven C1 grant ZKD1118 C16/16/005. KP is supported in part by the U.S. Department of Energy grant DE-SC0011941.

\appendix
\section{Consistent truncation for AdS vacua}\label{App:chain}
Consider a Lagrangian of the following form in $D$-dimensional spacetime:
\begin{equation}
    S=\int\sqrt{g}\left[R -\frac{1}{2}(\partial\phi)^2- \frac{1}{2}{\frac{1}{p!}}\rme^{a\phi}F_p^2\right]\,.\label{eq:StartingAct}
\end{equation}
When compactifying this theory on a round $S^n$ it is consistent to truncate to $\phi$ and the volume scalar $\varphi$ only. The reduction Ansatz for the metric is then: 
	\begin{equation}
	\rmd s^2 = \rme^{2\alpha\varphi}\rmd s^2_{d} + \rme^{2\beta\varphi} \rmd s^2_n\,,
	\end{equation}
where $D=d+n$ and $\rmd s^2_n$ is the metric on the round $S^n$ of normalised curvature. To get $d$-dimensional Einstein frame gravity with a canonically normalised volume scalar we take:
\begin{equation}
\beta = -\frac{d-2}{n}\alpha\,,\qquad \alpha^2 = \frac{n}{2(d+n-2)(d-2)}\,.
\end{equation}
When $n=p=d$, we can have both magnetic flux threading the $S^n$ and electric flux filling non-compact spacetime. The resulting  effective potential then reads: 	
\begin{align}
 V(\varphi, \phi) =  \frac{1}{2}Q_E^2\rme^{-a\phi + \left(\alpha d -\beta n\right)\varphi}  +  \frac{1}{2}Q_M^2\rme^{a\phi+ \left(\alpha d -\beta n\right)\varphi} - R_n \rme^{\left(\alpha d +n\beta-2\beta \right)\varphi}\,.	
\end{align}
Whenever the product $Q_EQ_M$ is non-zero this potential allows an extremum describing an $\AdS_d\times S^d$ solution. 

For type II SUGRA we have various $p$-form field strengths coupled to the dilaton. Clearly only when $p=5$ we do obey the above condition, and this explains the origin of the $\AdS_5\times S^5$ solution, where the self-duality anyways enforce that there is both magnetic and electric $F_5$ flux. In this particular case, there is no dilaton coupling. However, for the other field strengths, we naively do not find a vacuum since $p\neq n$. We can remedy this by reducing on a torus first so that we go from 10 to $D$ dimensions in which the $p$-form field strength obeys $2p=D$. The ``dilaton" $\phi$ will then be a linear combination of the string dilaton and the torus volume. We do not write out the details here since it is straightforward, but one ends up with the following chain of vacua \eqref{chain}.

\section{More general \texorpdfstring{$\AdS_1\times S^1\times\mathbb{T}^8$}{} solutions}\label{app:gen}
Let us consider a modification to the Ansatz in Sec. \ref{sec:T8sol} where we maintain the same $F_1$ and $F_9$ fluxes in the geometry but we introduce all possible directions for the flux $ F_5=(1-i\star)\mathcal{F}$ consistent with the SUSY intersection in (\ref{eq:inter}),
\be
\begin{aligned}
\mathcal{F} =& \rmd \theta^{1234}\wedge p_1+\rmd \theta^{1256}\wedge p_2+\rmd \theta^{1278}\wedge p_3.
\label{eq:F5 partial}    
\end{aligned}
\ee
We introduced the notation:
\be
p_a= \gamma_a \rmd x + \delta_a \rmd y,\quad \rmd\theta^{ijkl}=\rmd\theta^i\wedge\rmd\theta^j\wedge\rmd\theta^k\wedge\rmd\theta^l
\ee
for $a=1,\,2,\,3$. We need to quantise the numbers $\alpha, \beta, \gamma_a, \delta_a$, which will involve all radii $L_x$, $L_y$, $L_i$. In this case, we get partial moduli stabilisation, as follows
\begin{align}
\int_{1234x} F_5 &= (2\pi)^5 \gamma_1\equiv (2\pi)^4(\alpha')^2 N_{3}^{(1)}\label{eq"1},\\
\int_{1256x} F_5 &= (2\pi)^5 \gamma_2\equiv (2\pi)^4(\alpha')^2 N_{3}^{(2)},\\
\int_{1278x} F_5 &= (2\pi)^5 \gamma_3\equiv (2\pi)^4(\alpha')^2 N_{3}^{(3)},\\
\int_{5678x} F_5 &= -\rmi(2\pi)^5 \tfrac{L_{5678}}{L_{1234}}\tfrac{L_x}{L_y}\delta_1\equiv (2\pi)^4(\alpha')^2 \tilde{N}_{3}^{(1)},\\
\int_{3456x} F_5 &= -\rmi(2\pi)^5 \tfrac{L_{3456}}{L_{1278}}\tfrac{L_x}{L_y}\delta_3\equiv (2\pi)^4(\alpha')^2 \tilde{N}_{3}^{(2)},\\
\int_{3478x} F_5 &= -\rmi(2\pi)^5 \tfrac{L_{3478}}{L_{1256}}\tfrac{L_x}{L_y}\delta_2\equiv (2\pi)^4(\alpha')^2 \tilde{N}_{3}^{(3)}.\label{eq"3}
\end{align}
where $L_{ijkl}=L_iL_jL_kL_l$.

The equations of motion are satisfied as we found in Sec \ref{sec:The solution}, except for the Einstein equations (\ref{constraint2} - \ref{constraint5}) which introduce new constrains as follows. $R_{\theta_i\theta_i}=0$ leads to:
    \begin{align}
 L_x\delta_i&=\rmi L_y\gamma_i,\,\forall i.\label{eq:deltagamma}
 \end{align}
Lastly, $R_{xx},\,R_{yy},\,R_{xy}=0$ gives:
\begin{align}
\alpha\beta=&\rmi\qty(\tfrac{\gamma_1\delta_1}{L_{1234}^2}+\tfrac{\gamma_2\delta_2}{L_{1256}^2}+\tfrac{\gamma_3\delta_3}{L_{1278}^2}),\label{eq:alphaSq}
\end{align}
so if we choose $\alpha\in\mathbb{R}$, $\gamma_i\in\mathbb{I}$, and by (\ref{eq:deltagamma}) $\delta_i\in\mathbb{R}$. Notice that this combination is just $-2e^{2\phi}F_xF_y=(F_5^2)_{xy}$, which explains the length combination appearing in (\ref{eq:alphaSq}). Finally, $R_{\theta_i\theta_j}=0$ for $i\neq j$ is automatically satisfied.

\subsection{Supersymmetry}
We proceed as Sec. \ref{sec:SUSY part} for the Killing-Spinor equations using the more general Ansatz of the $F_5$ flux based on (\ref{eq:F5 partial}). 

The projector coming out of the dilatino and gravitino SUSY transformation remains as (\ref{eq:sing proj}) and (\ref{eq:sec proj1}), however the combination of $\Gamma$-matrices inside the operator $K$ is generalized to:
\begin{equation}
\begin{aligned}\label{eq:few proj}
    (-\rmi)^zK=&\tfrac{\delta_1}{L_{1234}}\qty(\Gamma_{\hat{\theta}^1\hat{\theta}^2\hat{\theta}^3\hat{\theta}^4}+\Gamma_{\hat{\theta}^5\hat{\theta}^6\hat{\theta}^7\hat{\theta}^8})+\tfrac{\delta_2}{L_{1256}}\qty(\Gamma_{\hat{\theta}^1\hat{\theta}^2\hat{\theta}^5\hat{\theta}^6}+\Gamma_{\hat{\theta}^3\hat{\theta}^4\hat{\theta}^7\hat{\theta}^8})\\
    &+\tfrac{\delta_3}{L_{1278}}\qty(\Gamma_{\hat{\theta}^1\hat{\theta}^2\hat{\theta}^7\hat{\theta}^8}+\Gamma_{\hat{\theta}^3\hat{\theta}^4\hat{\theta}^5\hat{\theta}^6}).
    \end{aligned}
\end{equation}
For the transformation in (\ref{eq:oldSUSY}) we pick $z=0$ in (\ref{eq:few proj}) and we proceed exactly as explained in Sec. \ref{sec:SUSY part} to find $K\epsilon=0$ for the negative chirality choice in (\ref{eq:chirCnd}), which indicates this solution does not require more projectors. As we explained before, it doesn't matter in what directions we include the $F_5$ fluxes, the KS equations remain exactly as (\ref{eq:KSeqold}), and the resulting spinor $\epsilon$ has the same ill behavior that we described originally.

This issue is resolved with the transformation in (\ref{eq:novelSUSY}) where we pick $z=1$ in (\ref{eq:few proj}) and a positive chirality in (\ref{eq:chirCnd}). We can apply (\ref{eq:selectProj1}) as well as the analogous projectors:
\begin{align}
\qty(1\pm\Gamma_{\hat{\theta}^1\hat{\theta}^2\hat{\theta}^5\hat{\theta}^6})\epsilon=0,\qquad \qty(1\mp\Gamma_{\hat{\theta}^1\hat{\theta}^2\hat{\theta}^7\hat{\theta}^8})\epsilon=0.
\end{align}
Imposing one of these two projectors implies the other if we assume the previous projectors (\ref{eq:sing proj} - \ref{eq:sec proj1}) and the chirality condition (\ref{eq:chirCnd}). So having more general fluxes than the solution presented in the main text would lead to $1/8$ BPS configurations of the form:
\begin{equation}
    \epsilon=\epsilon_0\,\exp(-\rmi\tfrac{\rme^\phi}{4}\qty[\tfrac{\beta}{L_y}\mp\qty(\tfrac{\delta_1}{L_{1234}}+\tfrac{\delta_2}{L_{1256}}-\tfrac{\delta_3}{L_{1278}})](L_xx+\rmi\,L_y y))
\end{equation}
for the spinor to be well-behaved under $2\pi$ rotation in $x$ we require,
\begin{equation}
    \tfrac{\rme^\phi L_x}{4}\qty[\tfrac{\beta}{L_y}\mp\qty(\tfrac{\delta_1}{L_{1234}}+\tfrac{\delta_2}{L_{1256}}-\tfrac{\delta_3}{L_{1278}})]=2\pi\mathbb{N}
\end{equation}
where $\beta$ and $\delta_i$ are related through (\ref{eq:few proj}), and as such we can recover constant spinor solutions for some part of the parameter space, as it happened in Sec. \ref{sec:SUSY part}. However, notice that even the most general solution with the SUSY preserving fluxes does not reproduce the number of supercharges in \cite{Billo:2021xzh}.

\section{\texorpdfstring{D$(-1)$/D$7$}{} brane solutions from T-duality}\label{app:wave}
We follow the notation of \cite{Massar:1999sb} in order to describe the D$0$/F$1$/D$8$ bound state solution in string frame. The metric is given by:
\begin{equation}
    \rmd s^2 = -H^{-\frac12}h^{-\frac32}\rmd t^2 + H^{\frac12}h^{-\frac12}\rmd z^2 + H^{-\frac12}h^{\frac12}(\rmd r^2 + r^2 \rmd \Omega^2_7)\,.
\label{eq:massar}\end{equation}
The other fields that are turned on are the dilaton, the RR gauge field $A_1$, the NSNS $B$-field $B_2$ and the Romans mass $M$:
\begin{align}
   &\rme^\phi = H^{-\frac54}h^{\frac14}\,,\label{eq:massar2} \\
   &B_2 = -\frac{3 k r^5 M z}{(k + r^6)^2} \rmd t \wedge \rmd r \,, \label{eq:Bcomb}
\end{align}
 where 
 \begin{equation}
 H = 1 + M z,\quad h = 1 + k/r^6.
 \end{equation}
This solution is written in the ``St\"uckelberg frame" where $B_2$ has eaten the RR 1-form $A_1$ and became massive. This solution has many non-trivial properties, such as its singularity structure. Since our aim is simply to T-dualise this solution along the time-direction, we will not go into details and instead refer to the original papers \cite{Massar:1999sb, Janssen:1999sa, Imamura:2001cr, Bergshoeff:2003sy}. The main observation is that the non-trivial B-field signals the presence of fundamental strings stretched along the $z$-direction, perpendicular to the D$8$ brane. 

In order to take the limit $M\to 0$, we should split the massive 2-form field into the separate $B$-field coupling to fundamental strings and the RR 2-form field strength for the D$0$. The result is to replace the combined \eqref{eq:Bcomb} to\footnote{This differs by a factor of 2 from \cite{Massar:1999sb}, but is consistent with \cite{Bergshoeff:2003sy}, which appears to be the correct one.}
 \begin{equation}\label{eq:massar3}
    B = h^{-1} \rmd t \wedge \rmd z,\quad A = H h^{-1} \rmd t.
\end{equation}
Let us now  T-dualize over time, using the standard Buscher rules \cite{Hassan:1999bv}. Clearly, T-duality over time is subtle and should really be understood only as a tool to generate SUGRA solutions. Although the solutions that are generated this way are not solutions to the usual IIA/IIB SUGRA theories, they are to so-named II* supergravities \cite{Hull:1998vg, Dijkgraaf:2016lym}. To cut a long story short; after Wick rotation of the T-dualised D$0$ brane one gets the D$(-1)$ solution of ``ordinary'' Euclidean IIB SUGRA. We apply the same logic to the full D$0$/F$1$/D$8$ bound state. In string frame we find, before Wick rotation to Euclidean IIB, the following solution:
\begin{align}
    &\rmd s^2 = - H^{\frac12}h^{\frac32}\rmd t^2 -2 H^{\frac12}h^{\frac12}\rmd t \rmd z + H^{-\frac12}h^{\frac12}(\rmd r^2 + r^2 \rmd \Omega^2_7)\label{eq:combineddual1}\\
     &  \rme^{\phi} = H^{-1}h,\quad B = 0,\\
     & F_1 = M \rmd t+ M h^{-1}\rmd z + H \partial_r h^{-1} \rmd r.\label{eq:combineddual2}
\end{align}
The solution in Euclidean signature is found by the usual replacement: $t=-i t_E$. 
Note that the metric has cross-terms, typical to gravitational waves, which are T-dual to fundamental strings. After Wick rotation this generates a complex metric. 

In Einstein frame, the metric becomes
\begin{equation}
       \rmd s^2 = - H h\rmd t^2 -2 H \rmd t \rmd z + (\rmd r^2 + r^2 \rmd \Omega^2_7).\label{eq:combineddual3}
\end{equation}
 When taking $M\to 0$ we find a D$(-1)$/W solution, where the D$(-1)$ is smeared along all directions but $t$. When $k\to 0$ we recover the $7$-brane solution in non-standard coordinates. For more standard coordinates we can change $t\to t-z$ when $k=0$. 
 For the general solution with non-zero $k,\,M$, a similar coordinate change $t\to t-z/h$, would lead to new cross-terms with $t$ and $r$ instead. Therefore, there will always be some imaginary component of the metric, consistent with our explanation around equation \eqref{F1F1}. Interestingly this complex metric seems to be ``allowable" close to the instantons, in the sense of Witten's criterion \cite{kontsevich2021wick, witten2021note} as explained next.
 Before we do so, note that the metric (\ref{eq:combineddual3}) suffers from other issues. Given that the directions orthogonal to the D$7$ brane are $z$ and $t$, we have not found a way to ``unsmear'' $h$ in the $z$ direction. One could be tempted to construct a near-horizon solution of the type $``\AdS_1 \times S^1"\times \mathbb{T}^8$ by smearing the D$(-1)$ over the D$7$ worldvolume and declaring that worldvolume to be a $\mathbb{T}^8$. However, the harmonic functions of D$(-1)$ don't depend on any other coordinate than $r$, which is the radial coordinate in the would-be $\mathbb{T}^8$-directions. Hence, if we smear it in that direction, it simply becomes a constant; it's smeared over all coordinates. We would like instead to have a nontrivial dependence on either $t$ or $z$.

\subsection{Allowable complex metric}
After T-dualising the D$0$/D$8$/F$1$ solution, we found a complex metric \eqref{eq:combineddual3}. Here we verify whether this metric passes Witten's criterion for complex metrics \cite{witten2021note}.

The criterion is based on the point-wise condition:
\begin{equation}
    \text{Re}(\sqrt{\text{det}g}g^{i_1j_1}\dots g^{i_{p+1}j_{p+1}}F_{i_1\dots i_{p+1}}F_{j_1\dots j_{p+1}})>0\label{eq:KScrit}
\end{equation}
for any real non-zero $F_{p+1}$, and for a particular metric $g_{ij}$ in Euclidean signature; independent of whether the $p$-form content on a particular model is complex. The criterion can be recast into a necessary condition \cite{witten2021note}
\begin{equation}
    \text{Re}\left(\frac{g_{ij}}{\sqrt{\det g}}\right)>0,
\end{equation}
and a sufficient condition
\begin{equation}
    \sum_{i=1}^{D}|\text{Arg}\lambda_i|<\pi,\label{eq:sumlambdai}
\end{equation}
where $\lambda_i$ are the eigenvalues in the metric diagonalization.

In our particular model, we may express (\ref{eq:combineddual3}) after Wick rotation as
\begin{equation}
\begin{aligned}
    \rmd s^2 &= H h\rmd t_E^2 +2 \rmi H \rmd t_E \rmd z + \rmd r^2 + r^2 \rmd \Omega^2_7\\
    &\equiv g^{(2)}_{ij}\rmd x^i\rmd x^j+ \rmd r^2 + r^2 \rmd \Omega^2_7.
\end{aligned}
\end{equation}
Our metric is allowable if and only if $k > r^6$, i.e. only close to the D-instantons. Necessity is automatically satisfied given that $\det g^{(2)}=H(z)^2>0$ and $H(z)h(r)>0,$ since $H>0$, $h>0$ by construction. For sufficiency, we determined the eigenvalues of $g^{(2)}$, which are purely real as long as $h>2$, i.e. $k > r^6$.

\bibliographystyle{JHEP}
\bibliography{refs}
\end{document}